\documentclass[doublecol]{epl2} 
\usepackage{subfigure, graphicx}
\newcommand{\quarterwidth}{4cm}

\newcommand{\halfwidth}{8cm}

\title{A formal treatment of generalized preferential attachment and its empirical validation}
\shorttitle{Measuring preferential attachment} %Insert here a short version of the title if it exceeds 70 characters

\author{Ama\c{c} Herda\v{g}delen\inst{1} \and Eser Ayg\"{u}n\inst{2} \and Haluk Bingol\inst{1}}
\institute{                    
  \inst{1} Bo\v{g}azi\c{c}i University, Department of Computer Engineering - P.K. 2 TR-34342 Bebek, Istanbul, TURKEY\\
  \inst{2} Istanbul Technical University, Department of Computer Engineering - Ayaza\v{g}a Kamp\"{u}s\"{u} TR-34469, Istanbul, Turkey}


\abstract{
Generalized preferential attachment is defined as the tendency of a vertex
to acquire new links in the future with respect to a particular vertex
property. Understanding which properties influence link acquisition tendency
(LAT) gives us a predictive power to estimate the future growth of
network and insight about the actual dynamics governing the complex
networks. In this study, we explore the effect of age and degree on LAT by analyzing data collected from a new complex-network growth
dataset. We found that LAT and degree of a vertex are linearly
correlated in accordance with previous studies. Interestingly, the relation
between LAT and age of a vertex is found to be in conflict with the known models of network growth. We identified three different periods in the network's lifetime where the relation between age and LAT is strongly positive, almost stationary and negative correspondingly.
}

\newcommand{\defn}[1]{\textit{#1}}
\newcommand{\figref}{Fig.}
\newcommand{\eqnref}{Eq.}

\begin{document}

\maketitle
\section{Introduction}
\label{sec:Introduction} One of the most profound discoveries in
complex-network studies was realizing that the structure and dynamics of
many real-world networks do not follow a completely random but rather
organized behavior. The power-law degree distribution observed in many complex networks has attracted a considerable attention because it is a significant deviation from random behavior  \cite{Barabasi1999, Newman2003}.  In this study, we focus on the dynamics that lead to power-law degree distributions.

In a dynamic complex network, there is a continuous creation of vertices and
formation of links between the vertices (vertex and link removal can be
included in this abstraction as well). For many networks, it is natural to view this process as
a competition between the vertices to acquire the newly formed links \cite{Pennock2002}. The
resulting degree distribution will be based on the link acquisition tendency
(LAT) values of individual vertices. It is interesting to ask which vertex
properties effect link acquisition tendency in which ways.

In this study, we aim to analyze the effect of some basic properties
such as age and degree of a vertex on the link acquisition tendency. Such an analysis requires growth data of networks with precise
time stamps of the vertex and link creations. As Newman \cite{Newman2001} states,
obtaining such dynamic data is difficult and most of the time the time
resolution is low. Although a limited number of studies
analyze the preferential attachment in several networks, their focus is on
degree related preferential attachment \cite{Barabasi2002,Camille2005,Jeong2003,Newman2001,Redner2004}. Furthermore,
many of these reported studies analyze dynamic networks obtained from
social collaboration and scientific citation data. Considering these limitations of the previous studies, we
decided to use a more generalized methodology that will allow us to analyze
not only the effect of degree on link acquisition but also of other vertex
properties. Instead of analyzing a previously published dataset, we decided
to utilize a new dataset with a high time resolution that will allow us
to analyze the preferential attachment in short time scales and comes from a
previously unexplored domain. 

\section{Methodology}
\label{sec:Methodology}
We assume that the network contains directed links and a
vertex is said to acquire a new link if a new link terminating at that
vertex is formed. We define the generalized preferential attachment as the tendency to acquire
new links with respect to vertex properties \cite{Camille2005}. Age and
degree of the vertices are the two properties we will discuss. It is possible to
formalize the notion of preferential attachment without referencing a
particular vertex property. 

Our analysis aims to measure LAT as a function of the vertex properties being investigated. It is based on data collected during an interval $[t_0,t_0 +\Delta t]$ of the network's lifetime. First, we construct a snapshot graph of the network at $t_0$, record the properties of vertices in that graph, and assume they do not change significantly during the analysis interval $[t_0 ,t_0 +\Delta t]$. We group the vertices having the same property values together and calculate the average number of new links that each group acquires during the interval. The average number of new links as a function of the vertex property value is a measure of the effect of having a specific property value on the link acquisition tendency. It is possible to view this process as calculating an histogram. We assign each vertex to a bin according to its property value at $t_0$ and record the number of new links accumulated for each bin during the analysis interval. By applying appropriate normalization measures, it is possible to formulize this measure as a probability function conditioned on the property value, as we will see below. 

Let $m$ be a generic vertex property (e.g. age, degree, etc.) taking one of the following values $M=\{m_1 ,m_2 ,...,m_q \}$ for each vertex. $P(m=m_i)$ is the probability that a vertex has property value $m_i$. This probability distribution is shortly represented as $P(m)$. Let event $L$ denote the acquisition of a new link by a vertex. $P(L)$ is the probability for a particular vertex to acquire a new link. By definition, without any a priori information, $P(L)=1/n$ where $n$ is the number of vertices. The conditional probability $P(m=m_i\vert L)$ is the probability of observing a vertex with property value $m_i$ at the termination point of a newly formed link. This probability distribution function is shortly represented as $P(m\vert L)$ for notational simplicity. Finally, the conditional probability $P(L\vert m=m_i)$ is the probability that a particular vertex will acquire the next link to be formed given that the property value of the vertex is $m_i$. It is a measure of the effect of property $m$ on link acquisition.

By applying the Bayes formula, we can calculate $P(L\vert m=m_i)$ as follows: 

\begin{equation}  \label{eq1}
P(L\vert m=m_i )=\frac{P(m=m_i \vert L)\cdot P(L)}{P(m=m_i)}
\end{equation}

This value gives us the link acquisition tendency as a function of the property $m$ and is a measure of the LAT. Unfortunately, it is not possible to calculate it directly from the data. But we can calculate estimates of measures on the right side of \eqnref~\ref{eq1} to estimate $P(L\vert m)$. Let $\Delta l_{total}$ be the total number of links acquired by all vertices during the interval $[t_{0},t_{0}+\Delta t]$. Let $\Delta l_{m_{i}}$ denote the total number of links acquired by the vertices with property value $m_{i}$ at $t_{0}$, during the interval $[t_{0},t_{0}+\Delta t]$. $\hat{P}(m|L)$ serves as estimation for $P(m|L)$. It is calculated as follows: 

\begin{equation}
\hat{P}(m=m_{i}|L)=\frac{\Delta l_{m_i}}{\Delta l_{total}}  \label{eq3}
\end{equation}

If we plug in the empirical estimates of the sample distribution of the property $m$ of the vertices at $t_{0}$, $\hat{P}(m)$, and $\hat{P}(m|L)$ into the right side of \eqnref~\ref{eq1}, we can obtain an estimate value for the link acquisition tendency as a function of property $m$.

An important point that is worth being noted is the assumption that the distribution ${P}(m)$ does not change during the interval $[t_0 ,t_0 +\Delta t].$ In reality, as time passes, vertex properties such as degree and age change. Previous studies acknowledge this problem and propose using relatively small $\Delta t$ values compared to the lifetime of
the network \cite{Barabasi2002,Camille2005,Jeong2003,Newman2001,Redner2004}. In order to avoid the same problem, we use small $\Delta t$ values and assume that $P(m)$ is stationary during $[t_0 ,t_0 +\Delta t]$. The effect of different $\Delta t$ values will also be investigated.

A closer examination of our methodology reveals important similarities between the proposed LAT measurement method and the method adopted in \cite{Newman2001}. Both methods employ a time-window size parameter which regulates the length of the analysis interval and assume that the underlying preferential attachment mechanism is time independent (at least during the analysis interval). One difference is that the final preferential attachment measures reported by \cite{Newman2001} are relative probability values and it is not possible to compare the results of different analyses (either in time or for different networks) without carrying out a normalization beforehand. The LAT measures reported in this study are normalized conditional probability distributions and hence they have straightforward interpretations. Another difference is that our methodology assumes the degree distribution does not change significantly and uses the distribution sampled at the beginning of the analysis interval ([$t_0, t_0+\Delta t$]) for all calculations regarding that interval while \cite{Newman2001} uses the exact distributions for each individual link acquisition incident. Since both methodologies already rely on the assumption that preferential attachment dynamics remain time-independent during the analysis, such a simplification in the calculations are quite justifiable and rewarding given the easier normalization techniques.

%For a through analysis of the generalized preferential attachment, the network growth data should contain precise time stamps of link and vertex creation events because all measurements are based on this information. Any problem with the ``quality'' of the data in that sense will cause degradation in the reliability of the results. 
\section{Validation}
%Before analyzing the new dataset, we would like to test our new method on synthetic networks built according to well known network growth models: The Barabasi Albert (BA) model and the Dorogovstev Mendes (DM) model \cite{Barabasi1999, Dorogovtsev2000b}. Since we know the exact dynamics behind the network growth in both models, we can compare our results to the expected ones and see whether our new method correctly captures the dynamics or not.
Before analyzing the new dataset, we would like to test our new method on a synthetic network built according to well known network growth model: The Barabasi Albert (BA) model \cite{Barabasi1999}. Since we know the exact dynamics behind the network growth in BA model, we can compare our results to the expected ones and see whether our new method correctly captures the dynamics or not.
\begin{figure}[htbp]
\centering
\subfigure[Degree related LAT of BA model.]
{
   \label{fig:lat_degree_ba} 
	\includegraphics[width=\halfwidth]{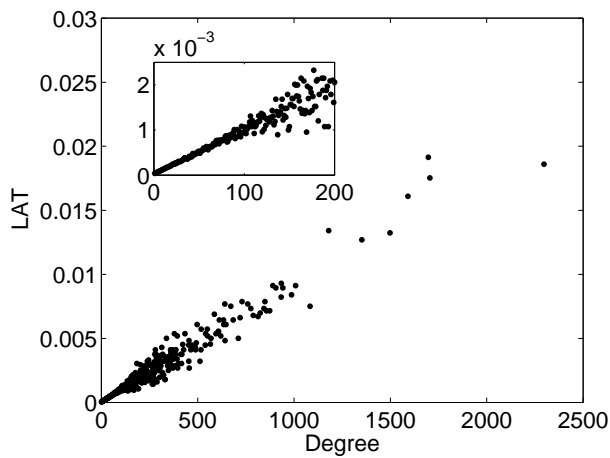}
}
\vfill
\subfigure[Age related LAT of BA model. The dashed line shows the analytically expected slope.]
{
	\label{fig:lat_age_ba} 
   \includegraphics[width=\halfwidth]{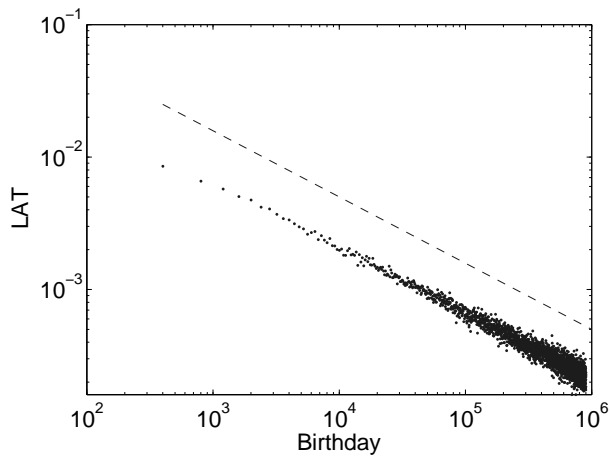}
}
\caption{LAT measurements for BA model.}
\label{fig:lat_validation}
\end{figure}

The BA network is created by setting the model parameters specified in \cite{Barabasi1999} as $t = 1,000,000$, $m_0=10$, and $m=6$. The final network contains approximately one million vertices and six million directed links. The first 900,000 vertices are used to construct the initial network and the remaining vertices are used to calculate the LAT measure. Our method correctly captures the linear degree based preferential attachment as seen in \figref~\ref{fig:lat_degree_ba}.  It is also analytically known that in the BA model the relation between the birthday of a vertex and the rate at which it increases its degree will be a power law ($LAT\propto m^{-0.5}$ where m is the creation time of a vertex) \cite{Barabasi1999}. Our method successfully captures the power law relation as shown in \figref~\ref{fig:lat_age_ba} along with a dashed line which has the analytically expected slope.
\section{Data}
In \cite{Herdagdelen2005}, a new network growth dataset which satisfies the aforementioned requirements is already introduced. The network is constructed by using the data crawled from ``Ek\c{s}i S\"{o}zl\"{u}k'' (literal translation from Turkish is Sour Dictionary) web site \cite{eksisozluk}. This site, which will be called the Dictionary shortly, is technically a collaborative hypertext dictionary in operation since 15 February 1999. The Dictionary is a site in which one can find explanations and definitions of almost any concept one can think of. Each concept is represented by a \defn{title}. Each individual definition about a title is called an \defn{entry}. The entries are listed chronologically under the titles and each entry has an associated timestamp indicating its time of creation. The entries may contain hyper-textual cross-references to other (possibly non-existing) titles and they have timestamps indicating the date and time they were written.

By using the raw data, we defined and constructed snapshot graphs of the Dictionary. A \defn{snapshot graph}, $G_t $, is a directed graph where each title $a$ is represented by a vertex $v_a $ and a cross-reference link from title $a$ to title $b$ is represented by an arc from vertex $v_a$ to vertex $v_b $. $G_t$ is constructed by including all vertices and links that were created until $t$. Several cross-references between the same titles are represented only once by a single arc in the graph. The time resolution of the creation times is one day for the first 2 years and one minute for the rest of the data.%So it is possible to obtain the snapshot graph of any given day (and even minute for most part) during the lifetime of the network.

In the end, we obtained a complex-network growth data which contains the vertex and link creation events dating back to the first day of the network (15 February 1999) extending until 01 January 2006. The final network has 1,921,425 vertices and 6,828,296 directed links. The degree distribution of the network follows a power law with the following form: $\hat{P}(m) \propto m^{-\gamma}$ where $\gamma=2.126$ \cite{Herdagdelen2005}.%The basic statistics of the final network on 01 January 2006 is given in \tabref \ref{tab:RawData}. The user number is the number of registered users at the time of crawling in the Dictionary.
%\scriptsize
%\begin{table}
%\caption{Characteristics of raw Dictionary Data.}
%\label{tab:RawData}
%\begin{center}
%\begin{tabular}{|l|l|l|l|}
%\hline
%Vertices & Links & From & To\\ 
%\hline
%1,921,425 & 6,828,296 & 1999-02-15 & 2005-12-31\\ \hline
%\end{tabular}%
%\end{center}
%\end{table}
%\normalsize

\section{LAT as a Function of Degree}

\label{sec:Results} 
%Our methodology depends on measurements made during individual intervals in the network's lifetime. Therefore, in order to draw general conclusions, we need to check whether our claims also hold for other intervals and are globally valid for the network. For this purpose we analyze single intervals for in-depth analysis and compare the results of several measurements in order to provide a global view of the network. In this section we let the generic vertex property $m$ represent first degree and then age of the nodes correspondingly.

Degree related preferential attachment is an important concept closely related to the degree distributions in networks. The linear preferential attachment hypothesis introduced by the Barab\'{a}si-Albert (BA) model states that the probability of a vertex to acquire new links is linearly proportional to its current degree \cite{Barabasi1999,Barabasi2002}. Almost all scale-free network models either explicitly incorporate the
linear preferential attachment hypothesis \cite{Barabasi1999,Krapivsky2000,Pennock2002} or expect it to
emerge from the interactions between the growth and dynamics of the network
\cite{Klemm2002,Vazquez2003}. There are some studies, which provide consistent results showing that there is indeed a linear preferential attachment phenomenon in some certain complex networks \cite{Barabasi2002,Camille2005,Eisenberg2003,Jeong2003,Newman2001,Redner2004}. %This common adoption is of no coincidence. It is analytically shown that for the BA model and its generalizations, a linear preferential attachment is necessary to obtain power-law degree distribution. For relations other than linear (i.e. sub linear or super linear) power-law degree distribution does not emerge \cite{Dorogovtsev2000,Krapivsky2000}. 

\begin{figure}[htbp]
\label{fig:lat_degree_figures}
\centering
\subfigure
{
	\label{fig:lat_degree}
	\includegraphics[width=\halfwidth]{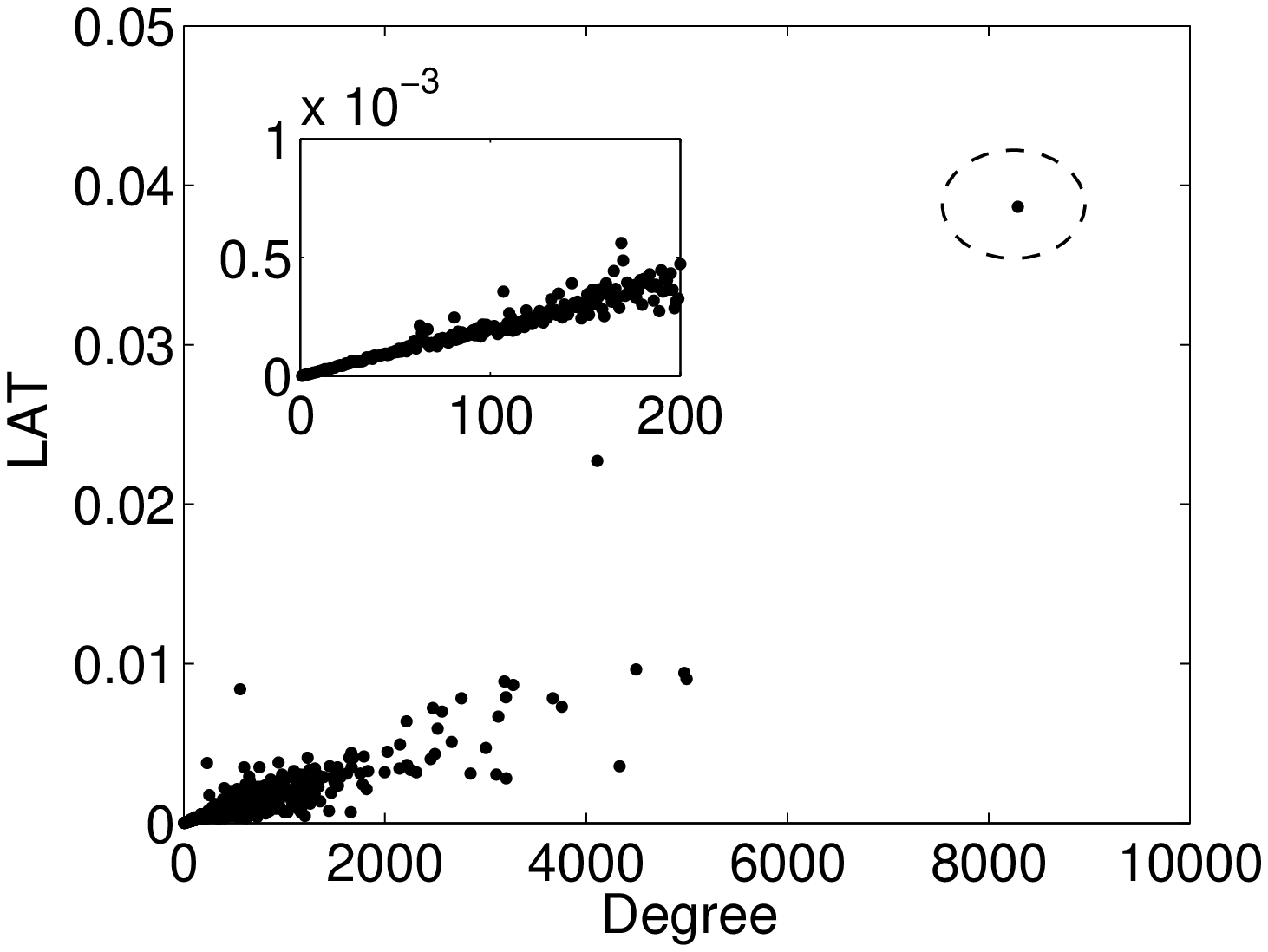}
}
\hfill
\subfigure
{
	\label{fig:lat_degree_newman}
	\includegraphics[width=\halfwidth]{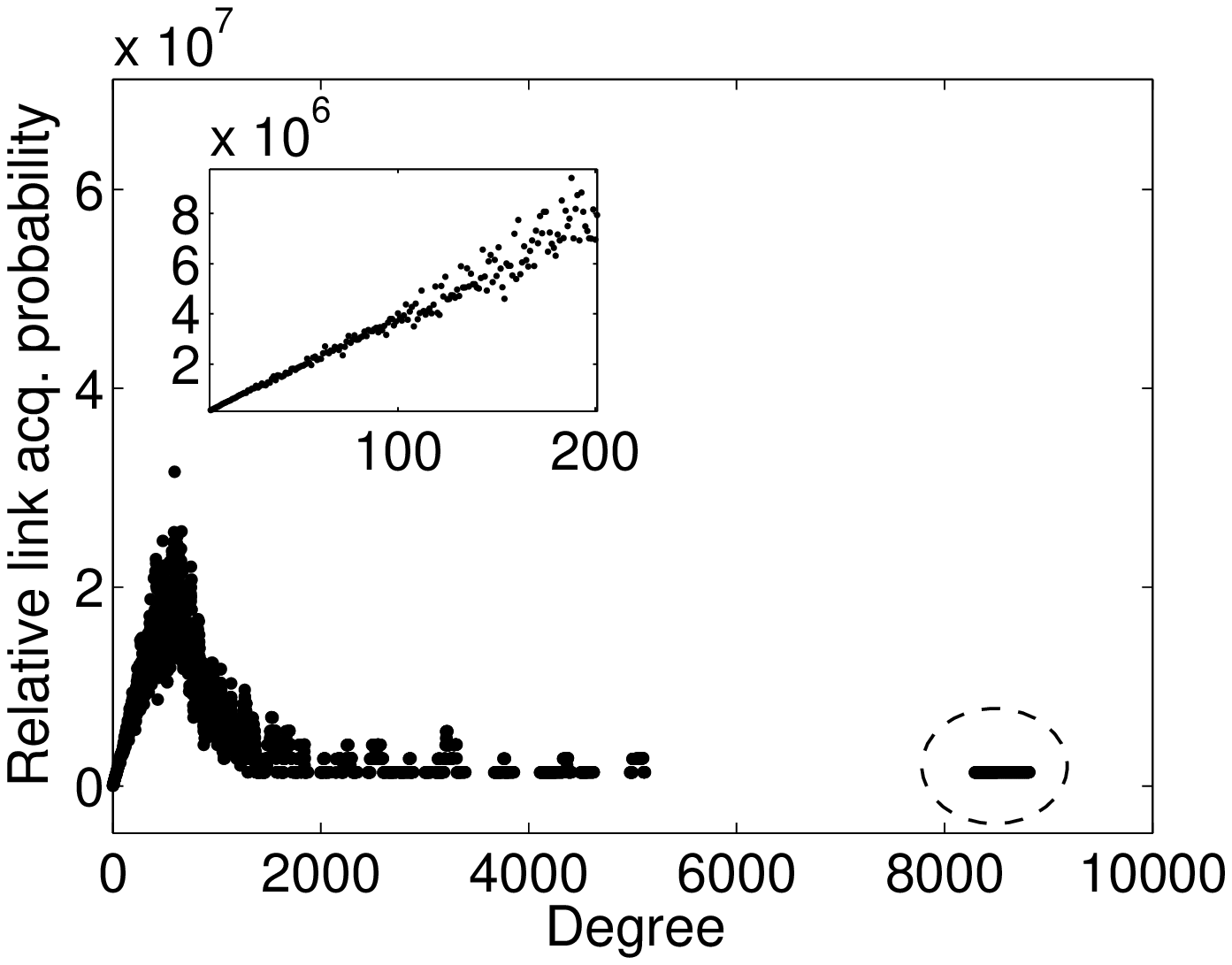}
}
\caption{\subref{fig:lat_degree} LAT vs. degree calculated by using  our newly proposed method and  \subref{fig:lat_degree_newman} preferential attachment measure obtained by the method adopted in \cite{Newman2001}, $t_0$= 01 December, 2005, $\Delta t= 31$ for both cases.}
\end{figure}

If we let  $m$ represent the degree of a vertex and construct the set $M=\{ m_1,m_2,...,m_q\}$ which is the set of all possible degree values in the network, then the LAT measure we calculate (i.e. $P(L\vert m)$) becomes the degree related preferential attachment.%The question of whether the preferential attachment is linear or not can be answered by testing if the LAT is linear (i.e. $P(L\vert m) \propto m$).
To measure degree related LAT, we constructed an initial network by using the snapshot of the network on 01 December 2005 and analyzed the network growth between 01 December 2005 and 31 December 2005. (i.e. $t_0$ is 01 December 2005, $\Delta t$ is 31 days). The calculated LAT values are plotted in \figref~\ref{fig:lat_degree}. The findings confirm that link acquisition tendency is linearly proportional to the vertex degree and the best fitting line has the form: $LAT=2.346.10^{-6}m+c$

We carried out the same analysis by using the method proposed by Newman in \cite{Newman2001}. Its results are plotted in \figref~\ref{fig:lat_degree_newman}. For lower values of the degree, the linearity is captured but for higher degree values the linear relation between the degree and preferential attachment measure disappears. We believe that in reality, the linearity exists even for the high degrees to some extent but the method fails to capture it. In \cite{Newman2001}, the bin to which a vertex is assigned shifts to the right for each link acquisition incident because the vertex's degree increases every time it acquires a new link. This factitiously leads to low preferential attachment values in the areas where the vertices are sparse. This effect is clearly  visible in the circled data points in \figref~\ref{fig:lat_degree} and \figref~\ref{fig:lat_degree_newman}. In both cases the data points are obtained for the same vertex. But in our methodology only one data point is produced for the vertex while in Newman's method a shifting series of data points are produced.

%\begin{figure}[htbp]
%\centering
%\subfigure[01 - 31 Jan 2002]
%{
%    \includegraphics[width=\quarterwidth]{lat_vs_degree_multi1.eps}
%}
%\hfill
%\subfigure[01 - 31 Dec 2002]
%{
%    \includegraphics[width=\quarterwidth]{LAT_vs_degree_multi2.eps}
%}
%\\
%\subfigure[01 - 31 Jan 2004]
%{
%    \includegraphics[width=\quarterwidth]{LAT_vs_degree_multi3.eps}
%}
%\hfill
%\subfigure[01 - 31 Dec 2004]
%{
%    \includegraphics[width=\quarterwidth]{LAT_vs_degree_multi4.eps}
%}
%\caption{Degree related LAT for different intervals.}
%\label{fig:multi_lat_degree}
%\end{figure}

\begin{figure}[htbp]
	\centering
	\includegraphics[width=\halfwidth]{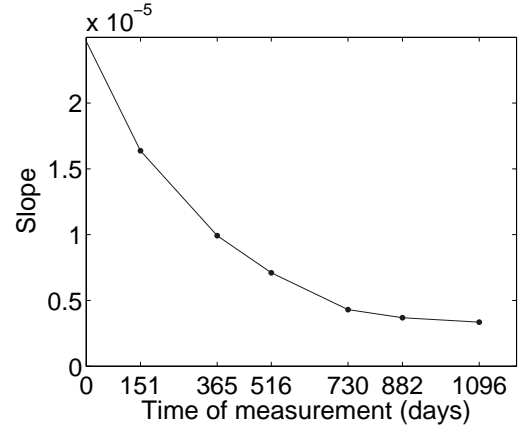}
	\caption{Change in the slope of degree related LAT value.}
	\label{fig:lat_degree_slope}
\end{figure}

In order to provide evidence that the observed linear dependence is a global property of the network and is not a temporary phenomenon specific to the interval 01 - 31 December 2005, we repeated measurements by using our original proposed method for different values of $t_0$ and obtained similar results for every interval we analyzed that is the relation between the LAT and degree of a vertex is linear independent of the time of the analysis. But the slope of the best fitting line changes significantly as time passes. Figure \ref{fig:lat_degree_slope} presents the slope values of the degree related LAT values which exhibit a significant decrease even for different months in the same calendar year.

\section{LAT as a Function of Age}
\defn{Age} of a vertex is defined as the number of days passed since the creation of the vertex. It is possible to consider the creation time (i.e. birthday) of a vertex in the analyses instead of its age and we do so in order to present our graphs in a compatible way with the previous studies \cite{Barabasi1999}. In this section, the generic property $m$ represents the birthday of a vertex in days.

%\figref \ref{fig:age_distribution} is plotted by using the sample $\hat{P}(m)$ distribution on 01 December 2005.

%\begin{figure}[htbp]
%	\centering
%	\label{fig:age_distribution}
%	\includegraphics[width=\fullwidth]{age_distribution.eps}
%	\caption{Age distribution on 01 December 2005.}
%\end{figure}

%The gaps seen in the graph (e.g. the one just before year 2001) correspond to the periods during which the Dictionary was closed due to maintenance.

\begin{figure}[htbp]
\centering
\includegraphics[width=\halfwidth]{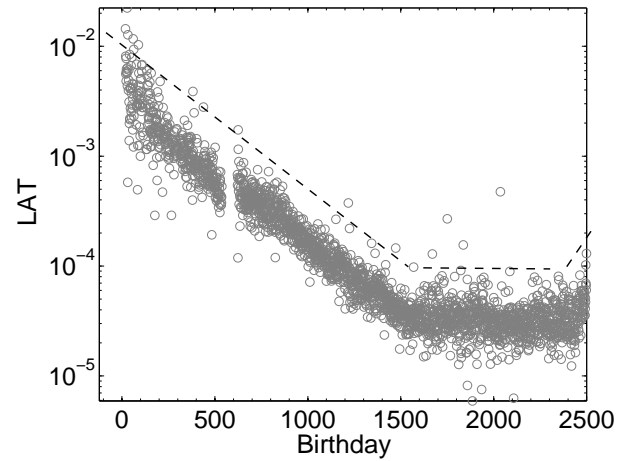}
\caption{LAT vs. age, $t_0 $= 01 December, 2005, $\Delta t=31$.}
\label{fig:lat_age}
\end{figure}

The age related LAT values calculated from the measurement done between 01 December 2005 and 31 December 2005 are given in \figref~\ref{fig:lat_age}. The creation times of the vertices are plotted in the x-axis in days with the first day corresponding to 15 February 1999. The small interval around approximately the $500^{th}$ day corresponds to period where the Dictionary was closed temporarily hence no data points exist for that interval.

Interestingly, the age related LAT does not follow a simple distribution. Instead, we identified three different vertex subsets according to their age values where the link acquisition tendency follows three different distributions. The subsets are named as \defn{old vertices}, \defn{middle vertices}, and \defn{young vertices}. The old vertices are the ones which are created approximately before March 2003 and have birthday values lower than 1000. For the old vertices, the relation between the birthday and link acquisition tendency of a vertex is strongly negative. Therefore, the earlier a vertex is created the higher LAT value it is expected to have. An exponential model provides a very good fit for LAT values for this period. The best fitting exponential model for the observation has the form $LAT\propto e^{-3\cdot 10^{-3}m}$.

Young vertices are the ones that are created during the last 60 days prior to the analysis. Among the young vertices, the relation between birthday and link acquisition tendency is positive which means being younger (i.e. having higher birthday values) pays off in terms of LAT value. An exponential model which has the form $LAT\propto e^{0.015m}$ is the best fitting exponential model.

The middle vertices are the ones that are created in between the old and young vertices. For the middle vertices, link acquisition tendency seems to be almost stationary with respect to the birthday. The best fitting line for the 
LAT values for this subset has the form $LAT=-4.583\cdot 10^{-9}m+c$ and it is almost a constant line for all practical purposes. 

We should stress that what we are after is not the precise boundaries between these sets of vertices. But the mere recognition of three different subsets of vertices according to their creation times suggests that the relation between link acquisition tendency and the age of a vertex adopts qualitatively different characteristics during the life time of the network. 

An analysis of variance (ANOVA) with a significance level of 0.05 confirmed that there is a main group effect of the birthday values: The mean LAT values for the three subsets differ significantly with old vertices being the highest, young vertices the second highest and the middle vertices the lowest.

\begin{figure}[htbp]
\centering
\subfigure[01 - 31 Jan 2002]
{
    \includegraphics[width=\quarterwidth]{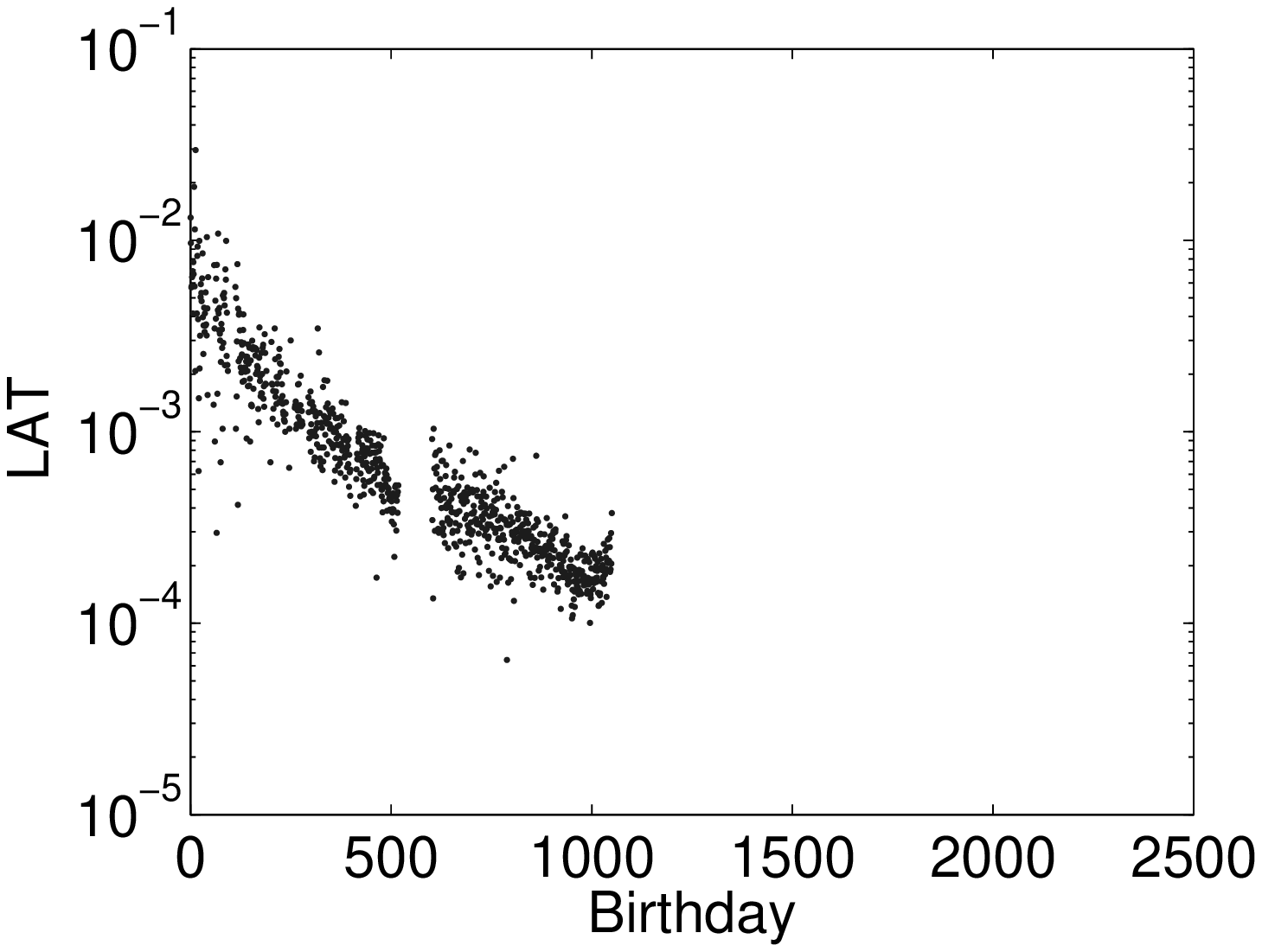}
}
\hfill
\subfigure[01 - 31 Dec 2002]
{
    \includegraphics[width=\quarterwidth]{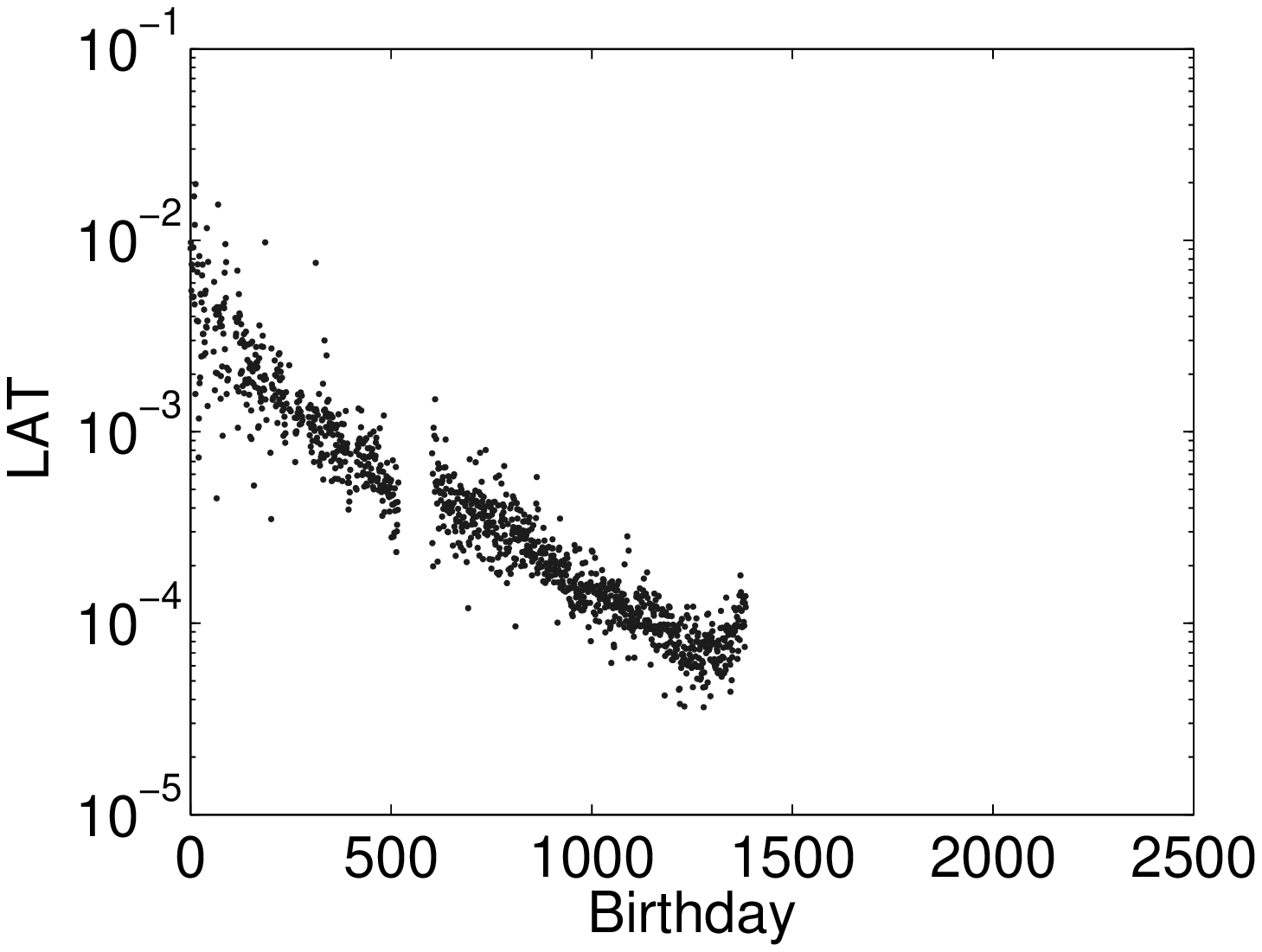}
}
\\
\subfigure[01 - 31 Jan 2004]
{
    \includegraphics[width=\quarterwidth]{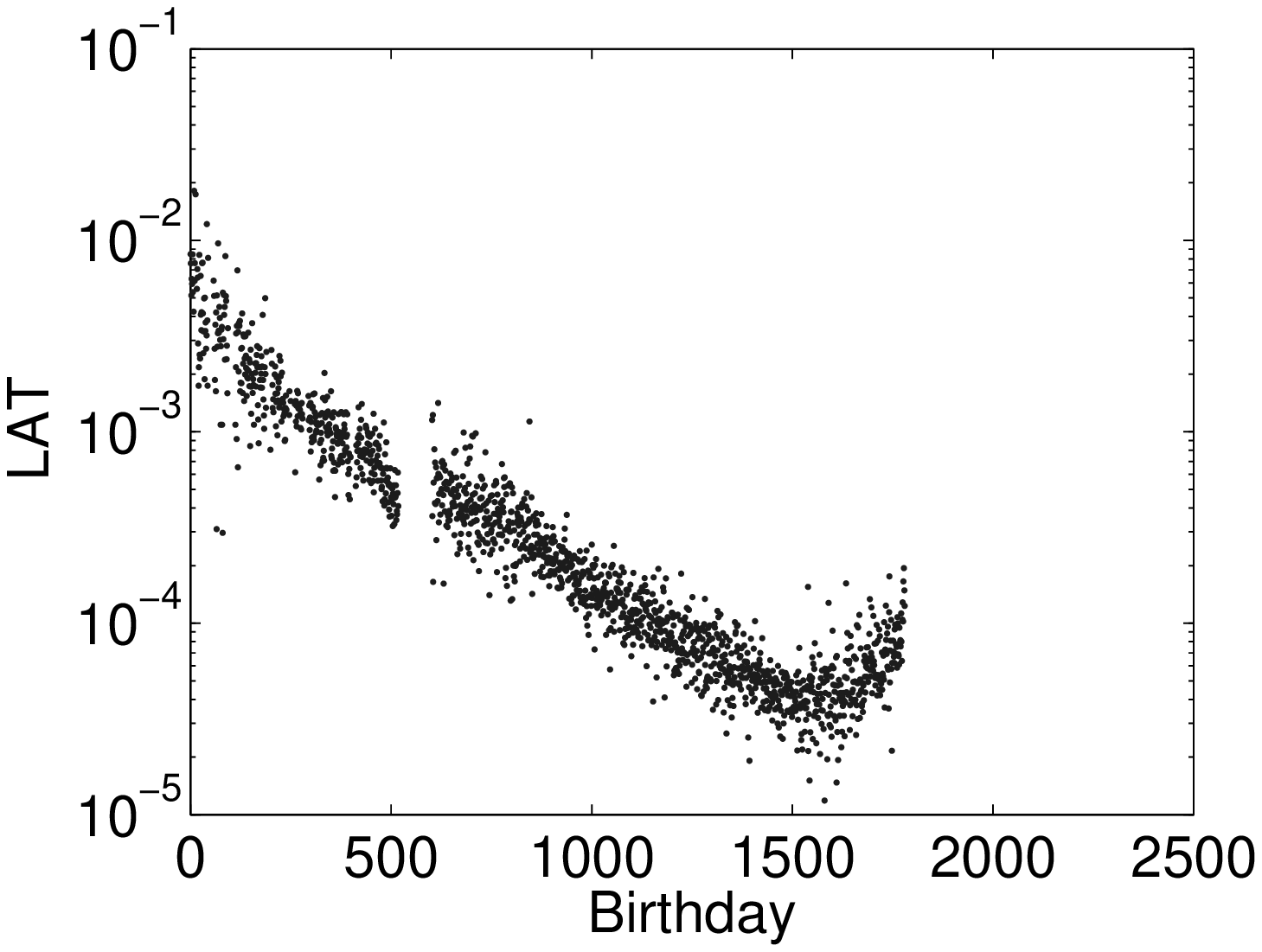}
}
\hfill
\subfigure[01 - 31 Dec 2004]
{
    \includegraphics[width=\quarterwidth]{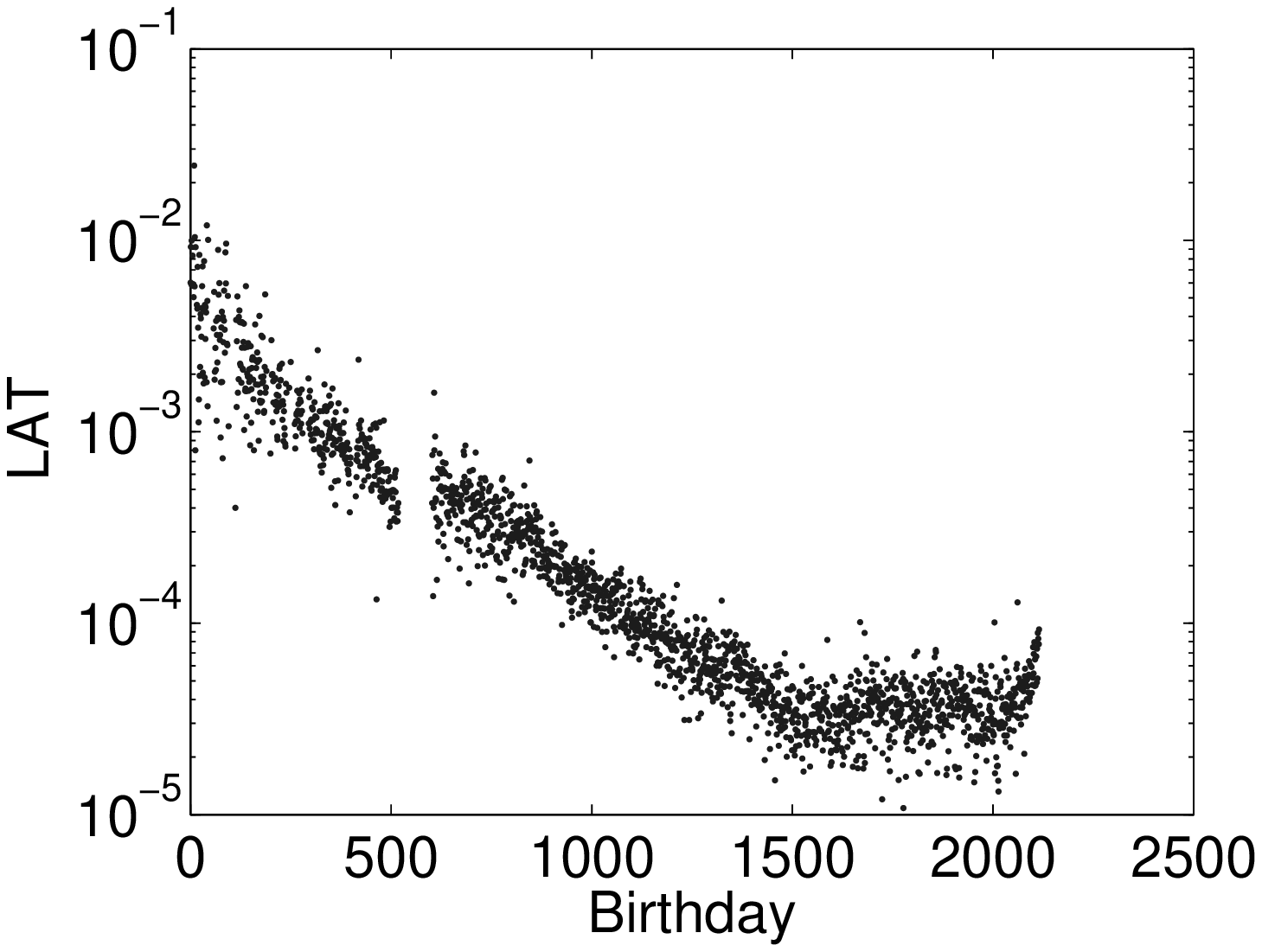}
}
\caption{Age related LAT for different intervals.}
\label{fig:multi_lat_age}
\end{figure}

In order to asses whether this partitioning of the lifetime of the Dictionary is valid not only for a single analysis but whole life time of the network, we carried out extensive measurements for different $t_0$ values and all of them yielded similar results. A sample of the age related LAT values calculated by analyzing different intervals are plotted in \figref \ref{fig:multi_lat_age}. Combining the results of ANOVA and the comparative analyses for different intervals, we conclude that partitioning the lifetime of the network into three periods with the aforementioned limits is globally valid for the network. The method adopted in \cite{Newman2001} also yields qualitatively similar results which confirm our findings but we are not presenting them here due to space limitations.

In order to analyze the individual characteristics of the three subsets of the vertices in more detail, we carried out 24 different measurements, spanning each calendar month starting from 01 January 2004 to 31 December 2005. For each interval, we calculated the correlation between the age and LAT values of old, middle, and young vertices separately. The average correlation between age and LAT for the old vertices is almost perfectly positive ($r=0.966$) and has a 95{\%} confidence interval (CI) of [0.964, 0.967]. The average correlation between age and LAT for the middle vertices is practically zero: $r=-0.005$ with a 95{\%} CI of [-0.060, 0.050]. The average correlation between age and LAT for the late period is strongly negative ($r=-0.528$) and has a 95{\%} CI of [-0.576, -0.479].

\section{The Effect of $\Delta t$}

\begin{figure}[htbp]
\centering
\subfigure[$\Delta t$ is 7 days]
{
    \includegraphics[width=\quarterwidth]{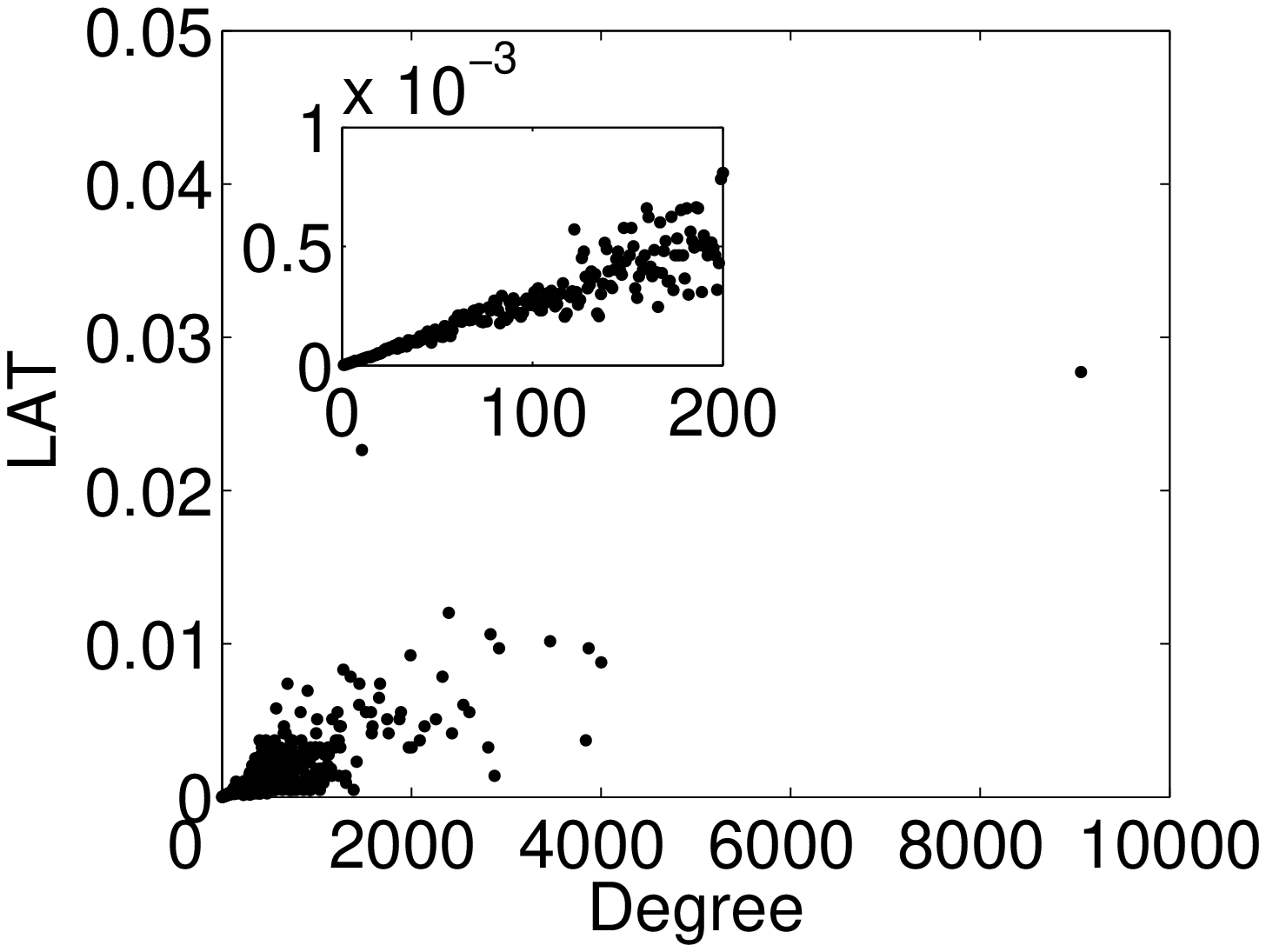}
}
\hfill
\subfigure[$\Delta t$ is 31 days]
{
    \includegraphics[width=\quarterwidth]{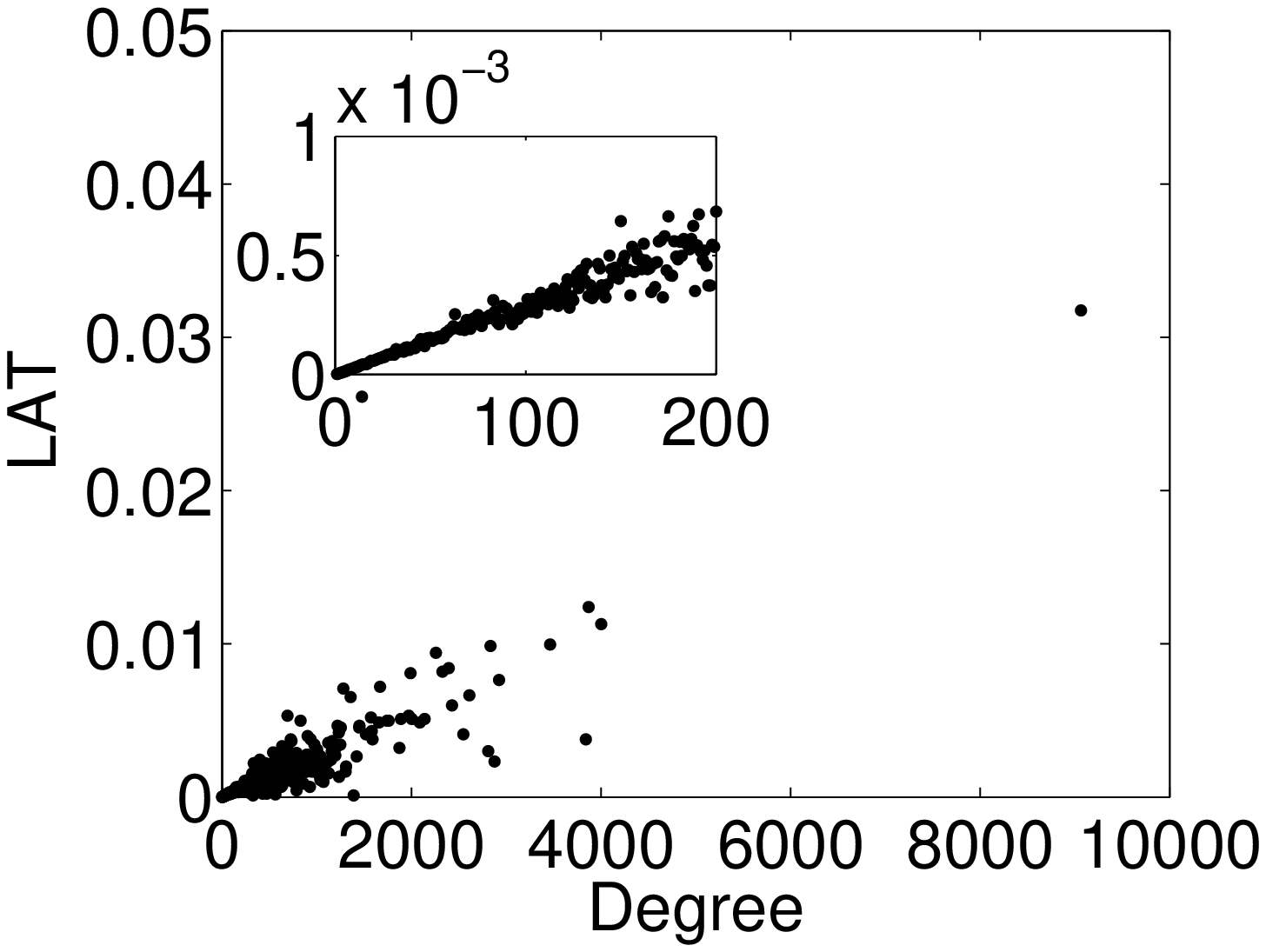}
}
\\
\subfigure[$\Delta t$ is 90 days]
{
    \includegraphics[width=\quarterwidth]{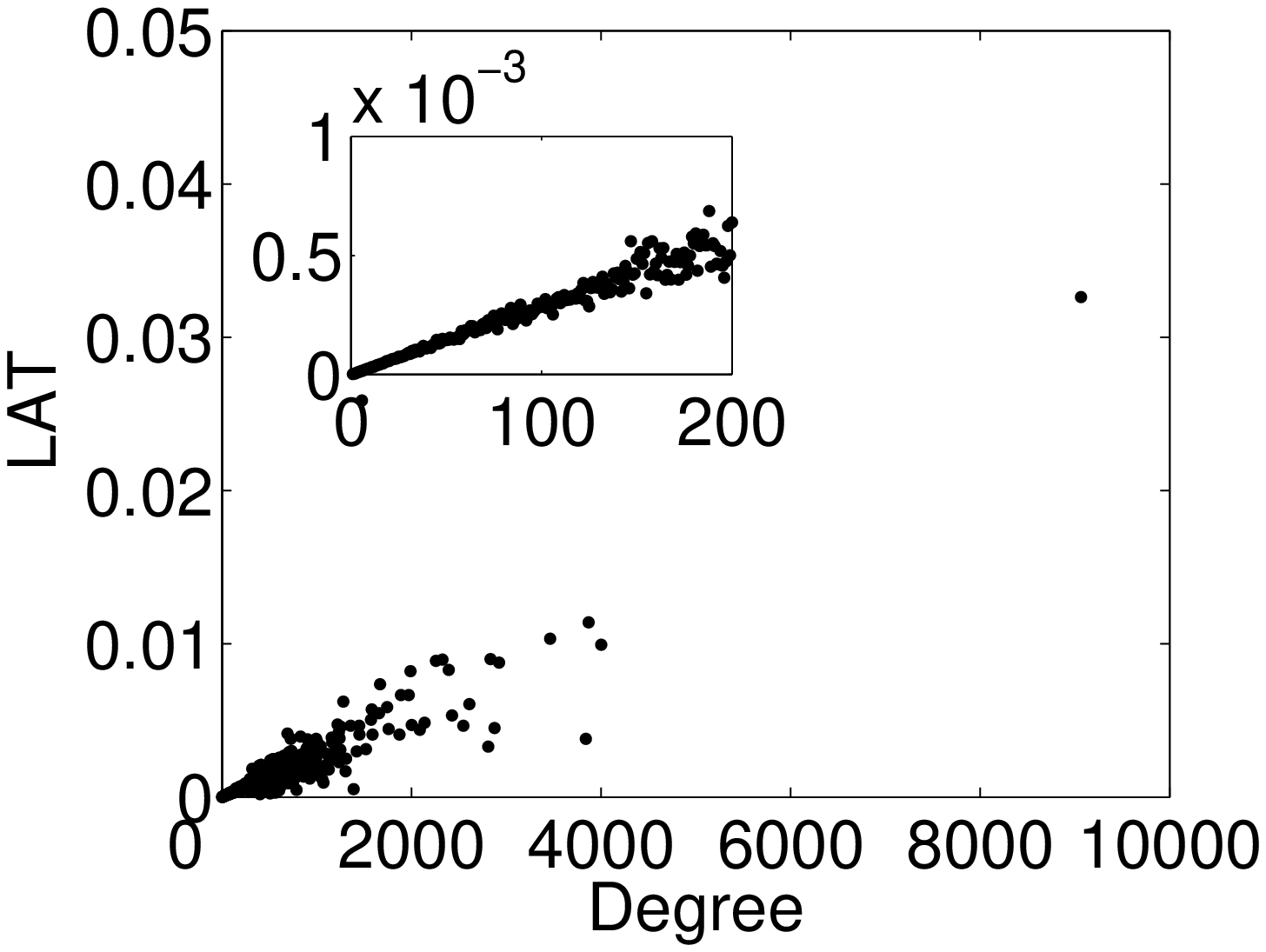}
}
\hfill
\subfigure[$\Delta t$ is 365 days]
{
    \includegraphics[width=\quarterwidth]{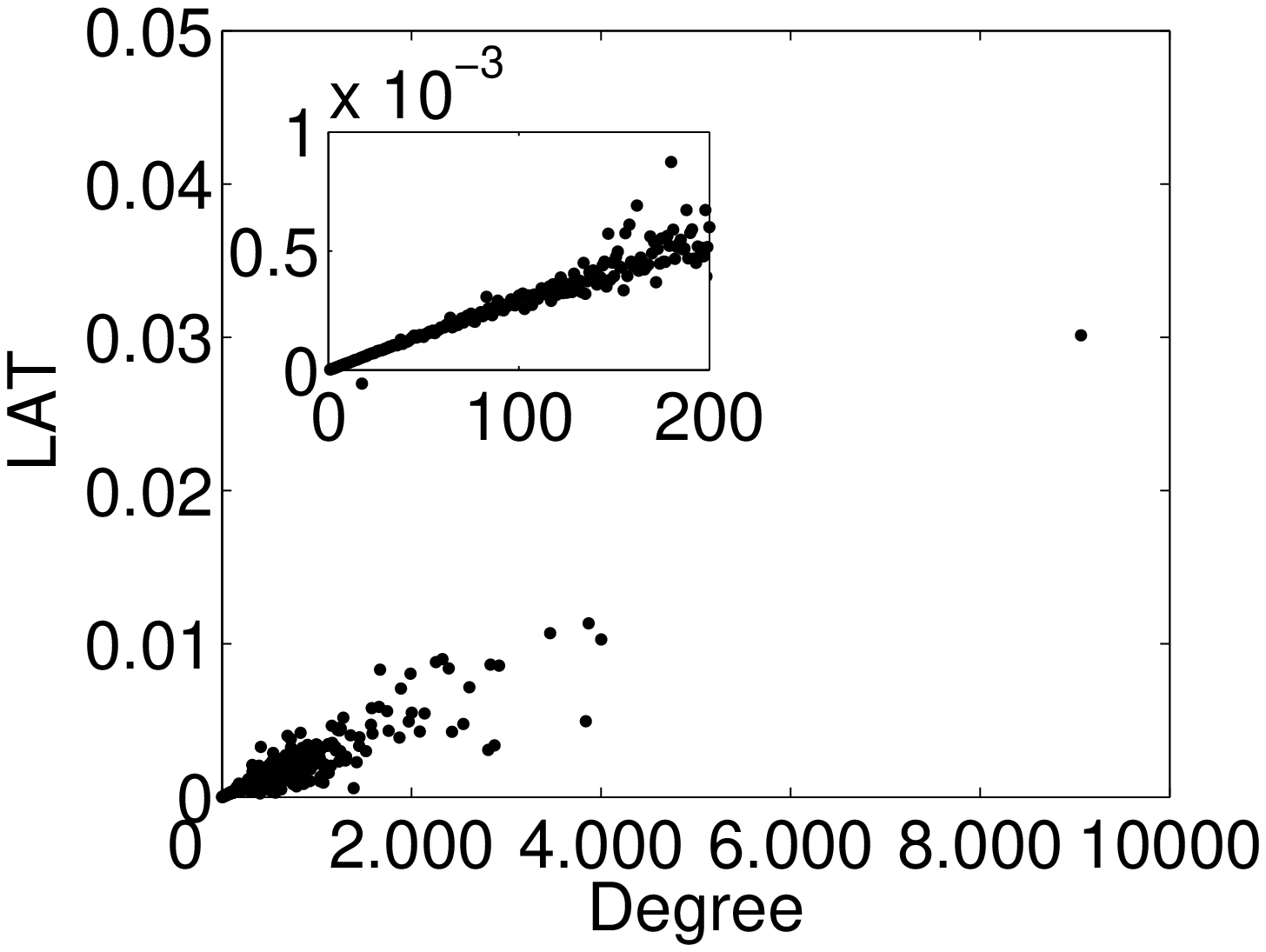}
}
\caption{Degree related LAT for different $\Delta t$ values ($t_0$ = 01 Jan 2004 for all cases).}
\label{fig:dt_lat_degree}
\end{figure}

\begin{figure}[htbp]
\centering
\subfigure[$\Delta t$ is 7 days]
{
    \includegraphics[width=\quarterwidth]{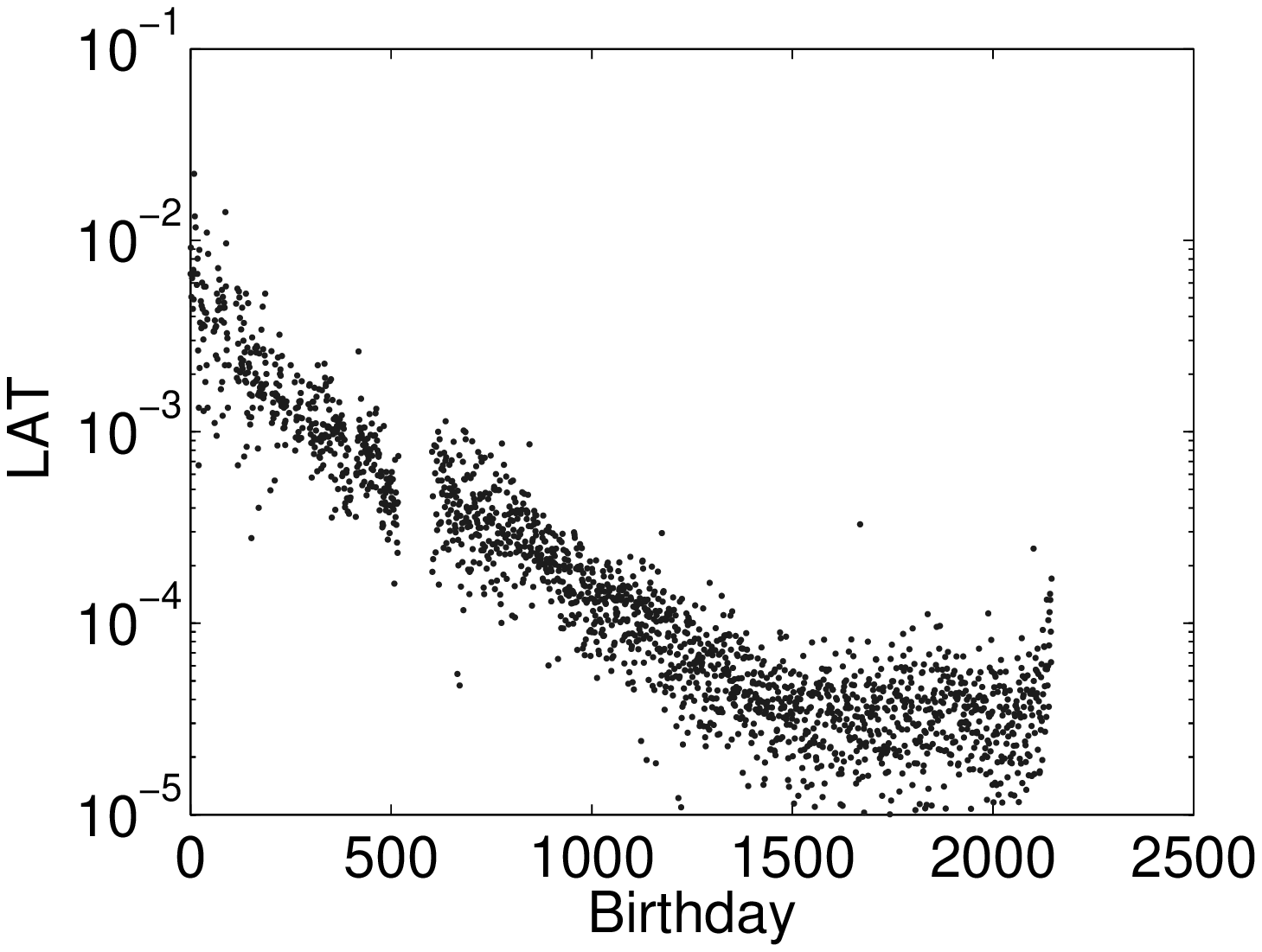}
}
\hfill
\subfigure[$\Delta t$ is 31 days]
{
    \includegraphics[width=\quarterwidth]{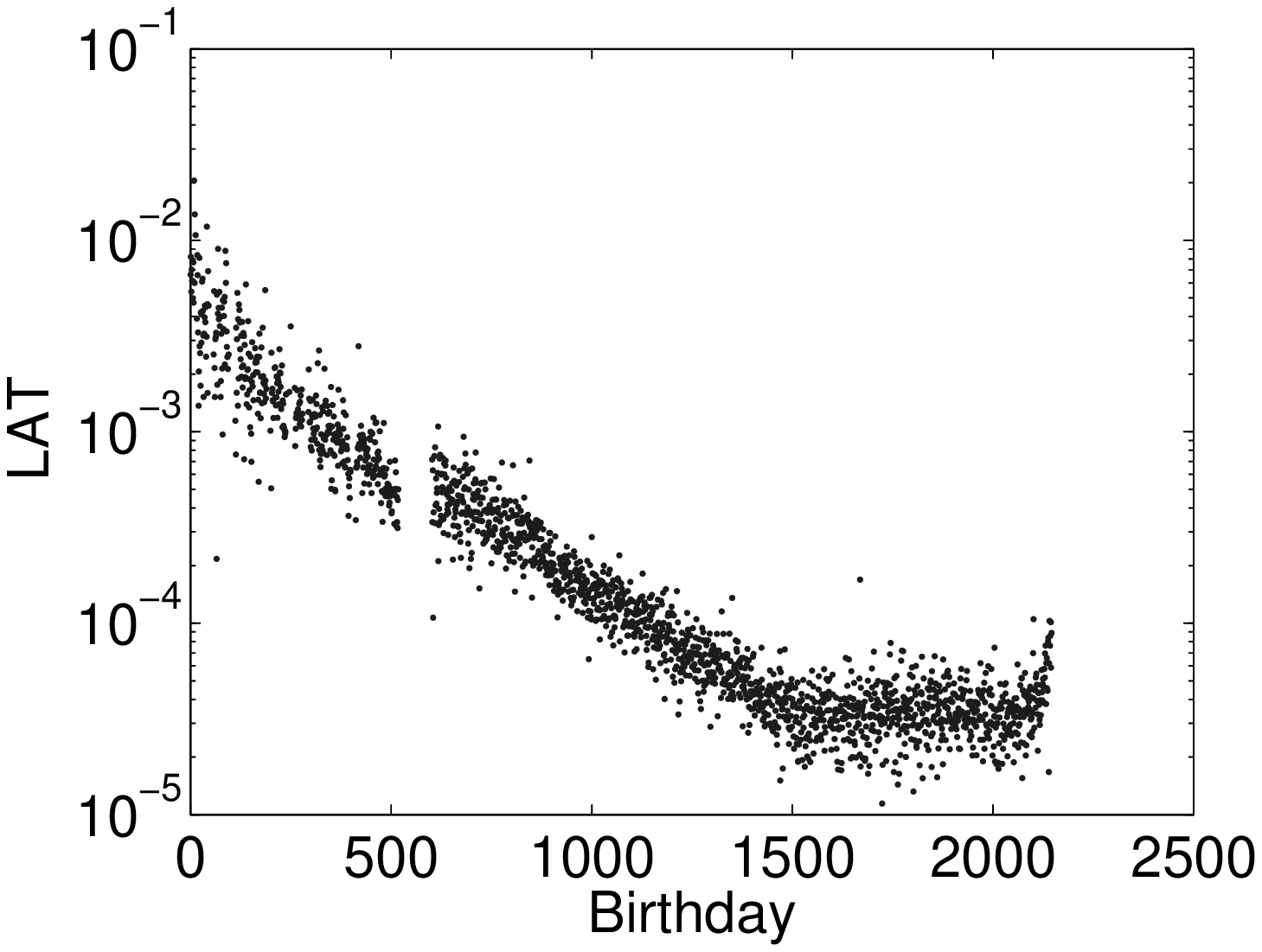}
}
\\
\subfigure[$\Delta t$ is 90 days]
{
    \includegraphics[width=\quarterwidth]{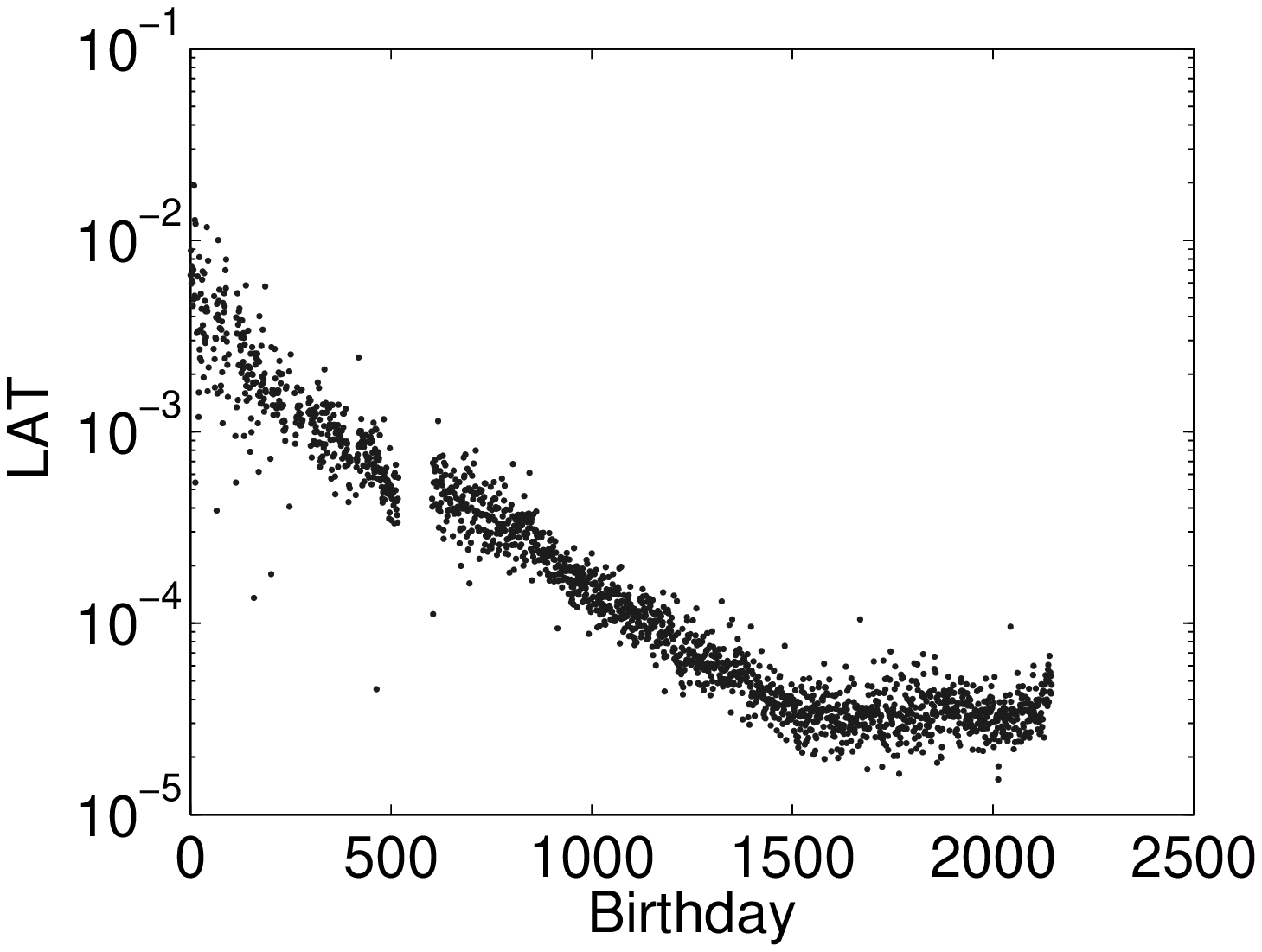}
}
\hfill
\subfigure[$\Delta t$ is 365 days]
{
    \includegraphics[width=\quarterwidth]{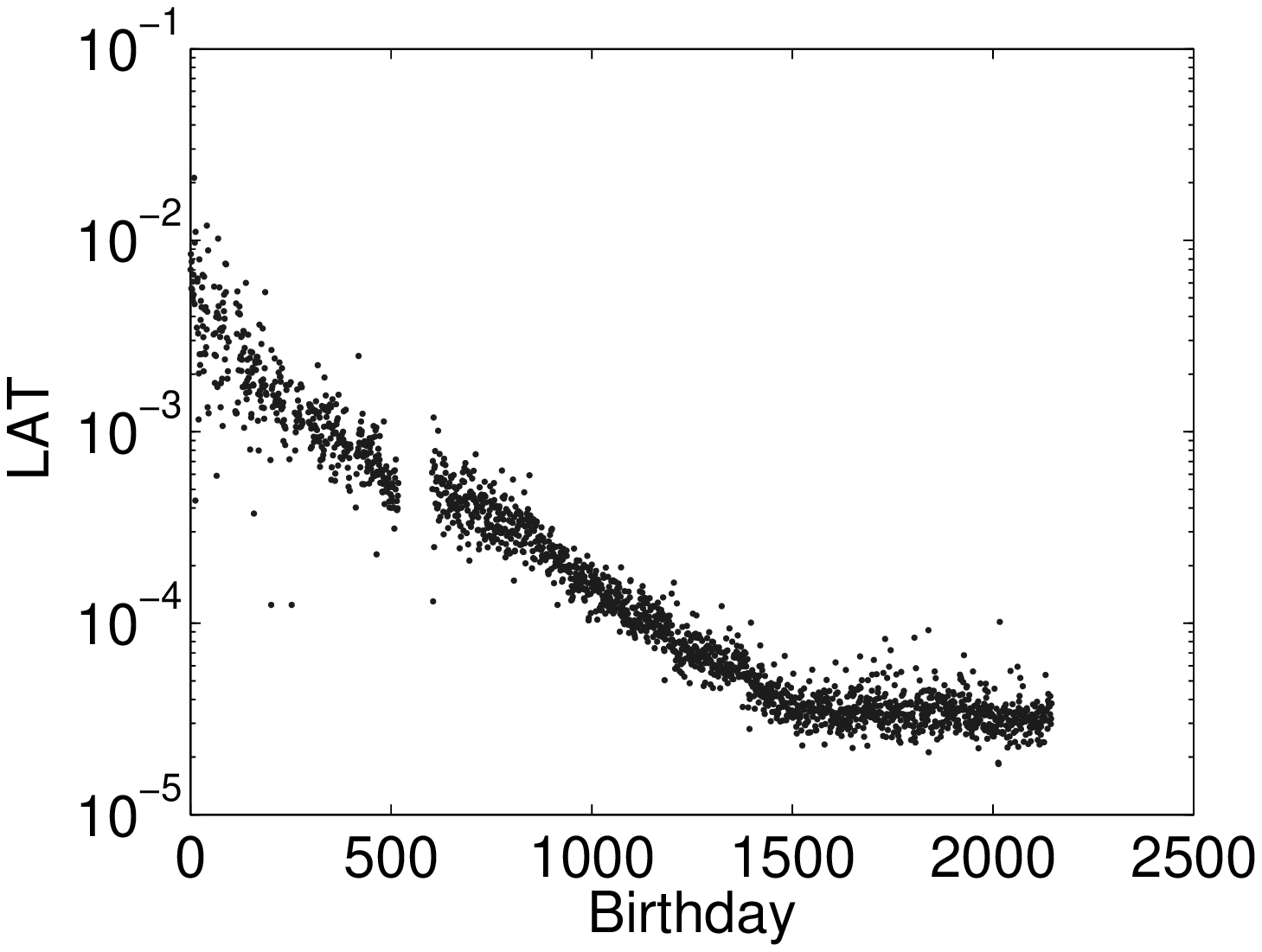}
}
\caption{Age related LAT for different $\Delta t$ values ($t_0$ = 01 Jan 2004 for all cases).}
\label{fig:dt_lat_age}
\end{figure}

The only free parameter in the proposed method is the length of the analysis window. We repeated the LAT calculations for degree and age for different values of $\Delta t$ to asses the importance of this parameter. A representative set of our calculations are plotted in \figref~\ref{fig:dt_lat_degree} and \figref~\ref{fig:dt_lat_age}. As seen in the figures, the degree related LAT does not change with respect to differing values of $\Delta t$. For the age related LAT value, however, the value of $\Delta t$ is more important. For longer $\Delta t$, we can not observe the increase in the LAT values of the young vertices. This is understandable because for longer time intervals our stationary $P(m)$ distribution assumption does not hold. The young vertices at the beginning of the analysis are no longer young at the end of the analysis when $\Delta t$ is one year and this interferes with the calculated LAT values.

\section{Conclusions and Future Work}

\label{sec:Discussion} 
As in compliance with previous studies, the relation between the LAT value
and degree of a vertex is found to be linear. The more connected vertices
are more likely to acquire new links in the future and this likeliness is a
linear function of vertex degree.

The relation between the LAT value and age of a vertex is more complicated. For the  old vertices that are created roughly before the $1500^{th}$ day (which corresponds to some time around March 2003) the LAT value is positively correlated with the age. For the vertices that are created after March 2003 (i.e. middle vertices), the relation between the LAT and age (or creation time) disappears except the vertices that are created during last 60 days prior to the analysis (i.e. young vertices). For the young vertices being younger pays off in terms of LAT.

In the real life dynamics, age of a vertex certainly does 
not play an explicit role in link acquisition. It is obvious that the users 
are not inclined towards giving cross references to other titles just 
because they are created early. This reasoning suggests that at least one 
mediator variable should be present and effecting LAT through age. The 
nature of this (or possibly these) variable(s) calls for future research. 
%Our candidate for such a mediator variable is \defn{conceptual elementariness}. When we say that concept $c_a $ is more elementary than concept $c_b $, we mean concept $c_a $ is more frequently used (i.e. cited, cross referenced) than concept $c_b $ is used when describing other concepts. Such a variable could shed some light on the reported differences between two sets of vertices and might bring an explanation why there is such a break down in March 2003. Obviously, our post-hoc explanation is only speculative given the lack of data concerning the ``conceptual elementariness''. A future study on this issue might involve collecting a conceptual elementariness measure (we admit that even providing a reasonable definition for conceptual elementariness is a challenging task, let alone providing a reliable measure for it), and analyzing the interaction between the age of a vertex and its conceptual elementariness.
\begin{acknowledgments}
This work was partially supported by Bogazici University Research Fund under the grant number 06A105. The authors thank Muhittin Mungan for stimulating discussions on the manuscript.
\end{acknowledgments}

\bibliographystyle{unsrt}

\end{document}